\documentclass[aps,pra,twocolumn,floatfix,showpacs,superscriptaddress]{revtex4}
\usepackage[dvips]{graphicx}
\usepackage{amssymb}
\usepackage{amsmath}

\begin{document}
\title{Ray-wave correspondence in lima\c{c}on-shaped semiconductor
  microcavities}
\author{Susumu Shinohara}
\affiliation{Max-Planck-Institut f\"ur Physik Komplexer Systeme,
  N\"othnitzer Stra\ss e 38, D-01187 Dresden, Germany}
\author{Martina Hentschel}
\affiliation{Max-Planck-Institut f\"ur Physik Komplexer Systeme,
  N\"othnitzer Stra\ss e 38, D-01187 Dresden, Germany}
\author{Jan Wiersig}
\affiliation{Institut f\"ur Theoretische Physik, Universit\"at
  Magdeburg, Postfach 4120, D-39016 Magdeburg, Germany}
\author{Takahiko Sasaki}
\affiliation{Center for Nano Materials and Technology, Japan Advanced
Institute of Science and Technology, 1-1 Asahidai, Nomi, Ishikawa
923-1292, Japan}
\author{Takahisa Harayama} \affiliation{NTT Communication Science
Laboratories, NTT Corporation, 2-4 Hikaridai, Seika-cho, Soraku-gun,
Kyoto 619-0237, Japan}
\newcommand{\etal}{{\it et al.}}
\begin{abstract}
The lima\c{c}on-shaped semiconductor microcavity is a ray-chaotic
cavity sustaining low-loss modes with mostly unidirectional emission
patterns.
Investigating these modes systematically, we show that the modes
correspond to ray description collectively, rather than individually.
In addition, we present experimental data on multimode lasing emission
patterns that show high unidirectionality and closely agree with the
ray description.
The origin of this agreement is well explained by the collective
correspondence mechanism.
\end{abstract}


\maketitle

Strong light confinement by whispering-gallery modes of microcavities
has attracted intense attention because of their potential
applications in the field of lasers and photonics \cite{Vahala03}.
Conventionally, the whispering-gallery modes are realized by
rotationally symmetric cavities such as disks \cite{McCall92} and
spheres \cite{Collot93}.
When applying these for laser cavities, however, one encounters two
undesired properties: low output power and the lack of emission
directionality, which are intrinsic to rotationally symmetric
cavities.
In order to resolve this problem, the idea of smoothly deforming a
cavity shape to a rotationally asymmetric one has been proposed
\cite{Noeckel94,Gmachl98}, which enables the existence of high quality
factor modes with highly anisotropic emissions.
N\"ockel et al. studied general influences of asymmetry on resonant
mode characteristics such as quality factors and emission patterns in
terms of the ray and wave chaos theory, revealing that microcavities
offer a good stage for the concepts and techniques of the classical
and quantum chaos theory to play an essential role \cite{Noeckel94}.
Since then, optical microcavities with various shapes have been
studied both theoretically and experimentally, with the motivation to
control light emission by a cavity shape and/or to understand the
effect of ray chaos on the cavity's wave characteristics
\cite{SB.Lee02, Hentschel02a, Wiersig03, Chern03,Kurdoglyan04,
Baryshnikov04, Fukushima04, Schwefel04, Fang05, Lebental06, SB.Lee07,
Tanaka07}.

One of the goals of using the cavity shape to control light emission
is the achievement of highly unidirectional microlasers.
Recently, on the basis of the ray and wave chaos theory, it has been
predicted that a semiconductor microcavity with a shape called the
lima\c{c}on of Pascal exhibits a highly unidirectional lasing emission
pattern, while maintaining relatively high quality factors \cite{Wiersig08}.
The lima\c{c}on shape is defined, in the polar coordinates
$(\rho,\phi)$, by $\rho(\phi)=R(1+\epsilon\cos\phi)$, where $R$ and
$\epsilon$ are size and deformation parameters, respectively.
The cavity shape is shown in the inset of Fig. 1 (a).
For $\epsilon$ values around 0.43, the ray dynamics inside the cavity
becomes predominantly chaotic.
The ray chaos results in a highly unidirectional emission, when the
refractive index $n$ is chosen around 3.3.
In Ref. \cite{Wiersig08}, it is reported that every low-loss resonant
mode exhibits a similar, or universal emission pattern corresponding
to a ray-calculated pattern, which is attributed to the ray-dynamical
emission mechanism governed by unstable manifolds \cite{Schwefel04}.
This analysis of resonant modes was carried out for the dimensionless
size parameter $nkR\approx 86$, where $k$ is the free-space wave
number.
Because low-loss modes, which are most likely to be excited in lasing
experiments, have similar highly unidirectional emission patterns,
laser light output to a single direction is expected with this cavity
shape.
This robust mechanism for the appearance of unidirectional lasing
emission differentiates the lima\c{c}on shape from the other shapes
previously proposed to achieve unidirectionality \cite{Chern03,
  Kurdoglyan04, Wiersig06}.
Motivated by such an attractive feature, Song et al. and Yan et
al. have fabricated lima\c{c}on-shaped semiconductor microcavities
independently \cite{Song08,Yan09}.
Although they both succeeded in observing mostly unidirectional
outputs, detailed correspondence between experimental and
ray-calculated emission patterns has not been studied.
It is one of the aims of this paper to study such correspondence.

\begin{figure}[t]
\centerline{\includegraphics[width=72mm]{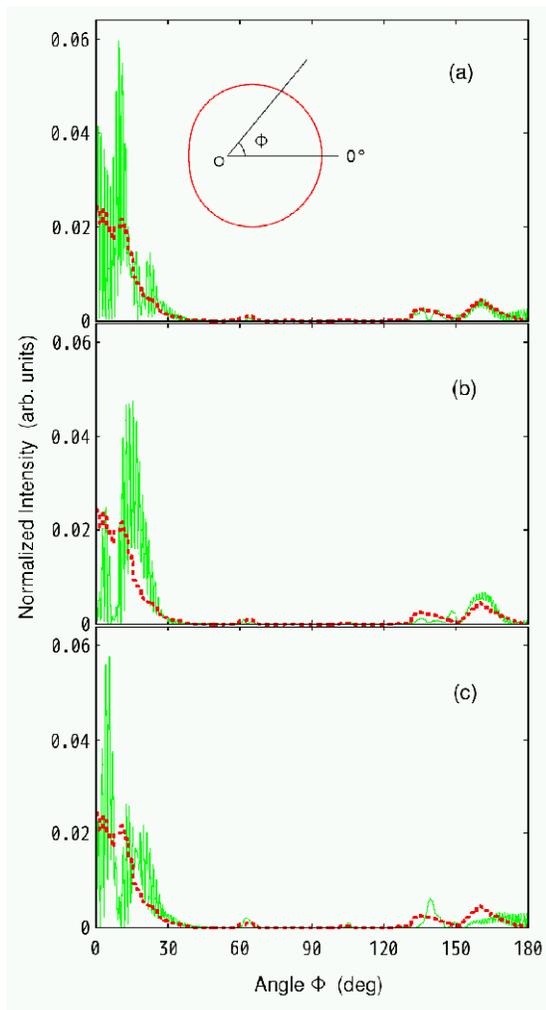}}
\caption{(Color online) Far-field emission patterns of the
lima\c{c}on-shaped microcavities. The solid curves are for the three
lowest-loss resonant modes with odd parity: (a)
$\mbox{Re}\,nkR=$484.11, $\mbox{Im}\,nkR=-$0.0017; (b)
$\mbox{Re}\,nkR=$483.84, $\mbox{Im}\,nkR=-$0.0017; (c)
$\mbox{Re}\,nkR=$484.75, $\mbox{Im}\,nkR=-$0.0024.  The dotted curve
is a ray-calculated pattern. The far-field angle $\phi$ is defined in
the inset of (a). Because the patterns are symmetric with respect to
the $\phi=0$ axis, only the half angular range is shown.}
\end{figure}

In this paper, we study theoretically and experimentally the
lima\c{c}on-shaped semiconductor microcavities larger than those
previously studied.
For larger cavities, better correspondence between experiment and
ray-calculation is expected, since ray description is generally
expected to work better in a shorter wavelength regime.
We theoretically investigate resonant modes for $nkR$ $\approx$ $484$
and perform experiments for fabricated lima\c{c}on-shaped
single-quantum-well laser diodes with $nkR$ $\approx$ $484$ and $nkR$
$\approx$ $1210$.
These sizes are much larger than the cavities of Song et al.  ($nkR$
$\approx$ $48$) and those of Yan et al. ($nkR$ $\approx$ $161$).
In the theoretical analysis, we point out a phenomenon revealed only
for larger $nkR$, that is, mode-dependent discrepancies of individual
modes' emission patterns from the ray-calculated pattern, which is
contrasted with the observation of the universal pattern for smaller
$nkR$ \cite{Wiersig08,SB.Lee07}.
In spite of such discrepancies, we demonstrate that ray-wave
correspondence can be recovered by averaging emission patterns of many
low-loss modes.
This averaged pattern can be considered as the approximate of a
multimode lasing emission pattern.
In experiments, we observe highly unidirectional lasing emission
patterns, which closely correspond to the ray-calculated pattern even
for smaller subpeaks.

\begin{figure}[t]
\centerline{\includegraphics[width=80mm]{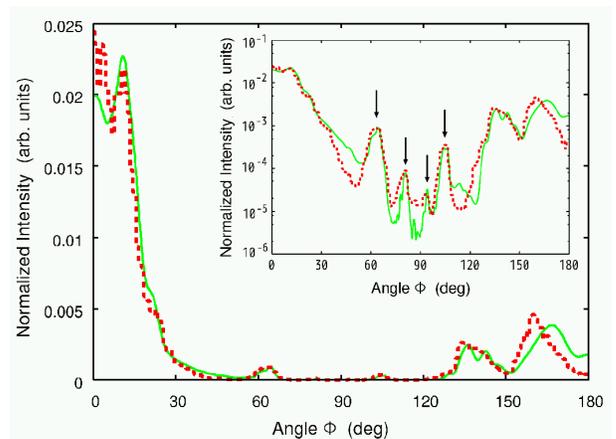}}
\caption{(Color online) The average of the far-field emission patterns
  for the 42 lowest-loss modes (solid curve) and the ray-calculated
  pattern (dotted curve). The inset shows the semi-log plot of these
  patterns.}
\end{figure}

First, we report theoretical results of resonant modes for $nkR\approx
484$.
Taking into account that experimentally observed light is transverse
electric (TE) polarized, we calculated resonant modes for the TE
polarization by using the boundary element method \cite{Wiersig03b}.
In Fig. 1, we plot far-field patterns of low-loss modes together with
the ray-calculated result.
We found 112 low-loss modes in the range $483.8<nkR<485.4$.
The modes shown in Fig. 1 are the three lowest-loss odd-parity modes.
The far-field patterns are normalized so that integration becomes
unity.
Each of these far-field patterns shows correspondence with the
ray-calculated pattern to some extent.
However, closer inspection reveals mode-dependent slight deviations
from the ray-calculated result that do not vanish even when rapid
oscillations are smeared out.
Such deviations have been observed also for resonant modes of the
stadium-shaped cavities with $nkR\approx 330$ \cite{Shinohara08,
Choi08}.
These results suggest that, in general, the emission pattern of an
individual mode does not converge to a ray-calculated pattern in the
short-wavelength limit, even after wave interference effects are
smeared out.
On the contrary, in smaller $nkR$ cases, an individual mode lacks
enough resolution to manifest its deviation from the ray-calculated
pattern, which appears to be a reason less attention was paid for the
deviations of individual modes from the ray calculation in the
previous works on smaller $nkR$ cases.

Next, we demonstrate that the deviations of individual modes are
averaged out, if the far-field patterns are averaged over many
low-loss modes.
In Fig. 2, we show the averaged far-field pattern for 42 lowest-loss
modes, where one can observe improved agreement between the wave and
ray calculations.
Although there is an arbitrariness on how we put a weight to each mode
when taking the average, here we put an equal weight.
Starting from a single mode, we checked that the agreement improves by
increasing the number of averaged modes $N$, and a converged pattern
is obtained for $N \approx 42$.
Notably, even tiny peaks at around $\phi=63, 80, 93, 104$ degrees of
the ray-calculated pattern can be reproduced in the averaged
wave-calculated pattern, which is clearly seen in the semi-log plot
shown in the inset of Fig. 2, where the four peaks are indicated by
arrows.
The convergence to ray calculation in the lima\c{c}on-shaped cavity is
better than that in the stadium-shaped cavity \cite{Shinohara08,
Choi08}.
We expect this difference can be explained by the fact that the
overlap of the Fresnel weighted unstable manifold with a leaky phase
space area for the lima\c{c}on-shaped cavity \cite{Wiersig08} is much
smaller than that for the stadium-shaped cavity \cite{Shinohara08},
whose detailed discussion will be presented elsewhere.

\begin{figure}[t]
\centerline{\includegraphics[width=75mm]{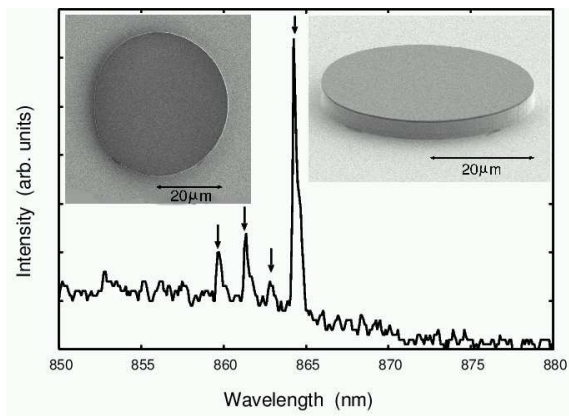}}
\caption{(Color online) Lasing spectrum for the lima\c{c}on-shaped
  single-quantum-well laser diodes with $\epsilon=0.43$ and
  $R=20\,\mu$m. The pumping current is 25 mA, which is slightly above
  the lasing threshold. Insets: Scanning electron microscope images of
  the lima\c{c}on-shaped semiconductor microcavity with
  $\epsilon=0.43$ and $R=20\,\mu$m before a contact layer is
  fabricated on the top: Top view (left inset) and oblique view (right
  inset).}
\end{figure}

\begin{figure}[t]
\centerline{\includegraphics[width=75mm]{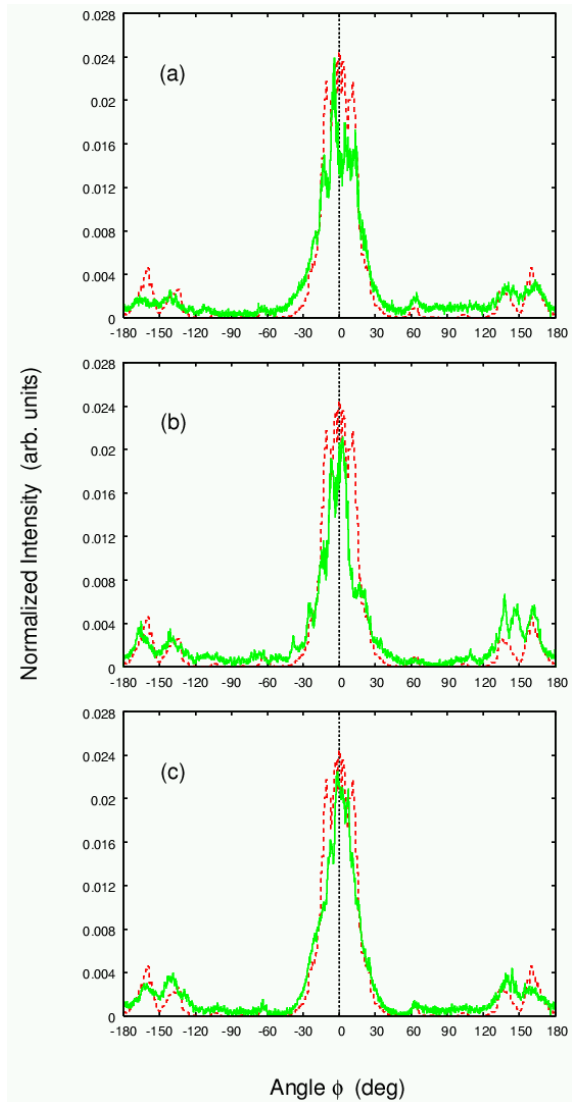}}
\caption{(Color online) Far-field emission patterns of the
  lima\c{c}on-shaped single-quantum-well laser diodes (solid curves)
  and the ray-calculated pattern (dotted curves). (a) and (b) are for
  two different samples with $R=20\,\mu$m, which are fabricated from
  the same epiwafer. (c) is for a cavity with $R=50\,\mu$m. The
  deformation parameter $\epsilon$ is 0.43 in all the cavities.}
\end{figure}

Concerning the rightmost peak at around $\phi$=$165$ degrees, one can
recognize a slight peak shift between the ray-calculated and
wave-calculated patterns.
By analyzing the ray dynamics, we found that, among all the tiny
peaks, only the rightmost peak is formed from rays refracted by the
cavity boundary with the incident angle close to the critical angle
for total internal reflection.
It is known that emission close to total internal reflection can be
largely affected by the Goos-H\"anchen effect and the Fresnel
filtering effect \cite{GH,Tureci02b}, causing discrepancies between
ray and wave calculations, although these effects are supposed to
vanish in the short-wavelength limit.
A detailed analysis is now in progress to identify the cause of the
peak shift, which will be reported elsewhere.

The above averaged pattern is not just an artificial product, but has
a physical meaning that it approximates a multimode lasing emission
pattern.
In order to study multimode lasing phenomena, one has to take into
account nonlinear interaction among modes caused by a lasing medium,
as has been done in Refs. \cite{Harayama03, Tureci07}.
Generally, there is no convenient method to predict properties of a
multimode lasing state from those of the (passive) modes involved in
lasing.
However, the averaged far-field pattern of low-loss modes with equal
weights can be considered as the time-averaged far-field pattern of a
multimode lasing state, provided that the gain band of a lasing medium
is so broad compared to an averaged mode spacing that many modes are
equally excited, and mode couplings are not so strong that a multimode
lasing state is dynamically described by a high dimensional torus.
If these assumptions are satisfied, one can expect that a multimode
lasing state is approximated by the averaged low-loss modes, thus
exhibiting an emission pattern closely corresponding to a
ray-calculated pattern.
Below, we examine this correspondence in the experiment of the
lima\c{c}on-shaped single-quantum-well laser diodes.

We fabricated lima\c{c}on-shaped microcavities with $\epsilon=0.43$
for $R=20\,\mu$m (i.e., $nkR\approx 484$) and $R=50\,\mu$m (i.e.,
$nkR\approx 1210$).
We note that the above wave calculations are directly applicable for
the fabricated cavities with $R=20\,\mu$m.
In the fabrication, we used a metal-organic chemical vapor deposition
grown gradient-index, separate-confinement-heterostructure,
single-quantum-well GaAs/Al$_x$Ga$_{1-x}$As structure.
The cavity geometry was defined by electron beam lithography and a
reactive ion etching technique.
The details of the structure and fabrication process are precisely
the same as those in Ref. \cite{Fukushima04}.
The effective refractive index of our microcavities is $n=3.3$, which
is estimated from the single-quantum-well epiwafer structure.
The insets of Fig. 3 show the scanning electron microscope images of
the lima\c{c}on-shaped microcavity with $R=20\,\mu$m, where one can
confirm the high degree of smoothness and verticality of the cavity
boundary.
Lasing experiments are carried out at room temperature and with
electrical pumping.
A pulsed current has 500-ns width at a 1 kHz repetition.
The lasing threshold is estimated as 22 mA for $R=20\,\mu$m and 34 mA
for $R=50\,\mu$m.
The lasing phenomenon was confirmed by the appearance of narrow peaks
in the spectrum, where the peaks are observed around 865 nm.
In Fig. 3, we plot lasing spectrum for the cavity with $R=20\,\mu$m
and pumping current 25 mA, which is slightly above the lasing
threshold.
The mode spacing of the four peaks in Fig. 3 is 1.7 nm, which
corresponds to the optical pathlength of the circumference, i.e.
$\triangle \lambda \approx \lambda^2/(2\pi n R)=1.8$ nm.
By increasing the pumping current, many modes are involved in lasing.
Although we could not resolve individual lasing modes because of the
resolution limit of our spectrometer, from the estimate of the
averaged mode spacing, we expect that at least 40 modes are involved
for the cavity with $R=20\,\mu$m at the current 200 mA.

Figure 4 (a)-(c) show measured far-field patterns (solid curves) with
the ray-calculated pattern (broken curves).
Figures 4 (a) and 4 (b) are for two different cavities with
$R=20\,\mu$m, which are fabricated from the same epiwafer, and Fig. 4
(c) is for a cavity with $R=50\,\mu$m.
The pumping current is $200$ mA for all the data in Fig. 4.
These experimental far-field patterns show close correspondence with
the ray-calculated patterns.
Remarkably, one can see correspondence even for smaller peaks at 136
and 160 degrees to some extent.
We checked that for sufficiently large pumping, such good
correspondence is observed robustly.

In summary, for the lima\c{c}on-shaped semiconductor microcavities, we
have shown that low-loss resonant modes exhibit mostly unidirectional
emission patterns, but there are mode-dependent slight discrepancies
from the ray-calculated pattern.
Nonetheless, we demonstrated that an averaged emission pattern of many
low-loss modes, which can be regarded as the approximate of a
multimode lasing emission pattern, closely corresponds to the
ray-calculated pattern.
In addition, we present experimental far-field data, which provide
convincing evidence not only of highly unidirectional lasing emission
but of the capability of the ray calculation to reproduce experimental
multimode lasing emission patterns.
We expect the results presented here can be applied to other cavities
having predominantly chaotic ray dynamics.

We would like to thank J.-W. Ryu and J. Unterhinninghofen for
discussions. S.S., M.H. and J.W. acknowledge financial support from
the DFG research group 760 ``Scattering Systems with Complex
Dynamics'' and the DFG Emmy Noether Program.

\end{document}